\documentclass[%
 reprint,
 amsmath,amssymb,
 aps,
 prx,
]{revtex4-2}

\usepackage{graphicx}% Include figure files
\usepackage{caption}% Accommodate subfigures
\usepackage{subcaption}%  "           "
\usepackage{dcolumn}% Align table columns on decimal point
\usepackage{bm}% bold math
\usepackage{hyperref}% add hypertext capabilities
\usepackage[mathlines]{lineno}% Enable numbering of text and display math
\usepackage{natbib}
\setcitestyle{super}

\usepackage[T1]{fontenc}    % teach bibtex to be more inclusive of other culture

\usepackage{tikz}
\usetikzlibrary{quantikz}   % draw quantum circuits

\usepackage{physics}        % lots of handy things, most importantly brakets

\newcommand{\bigw}[1]{$\Omega(#1)$}         % typing mathcal gets old...
\newcommand{\bigo}[1]{$O(#1)$}              %        big O
\newcommand{\bigq}[1]{$\Theta(#1)$}         %        big Θ

\graphicspath{{./}}      % specify directory to find graphics

\begin{document}

\preprint{APS/123-QED}

\title{A systematic variational approach to band theory in a quantum computer}

\author{Kyle Sherbert}
\author{Frank Cerasoli}
\author{Marco Buongiorno-Nardelli}
\affiliation{
 Department of Physics, University of North Texas, Denton, TX 76303, USA.\\E-mail: mbn@unt.edu
}

\date{\today}

\begin{abstract}
Quantum computers promise to revolutionize our ability to simulate molecules, and cloud-based hardware is becoming increasingly accessible to a wide body of researchers.
Algorithms such as Quantum Phase Estimation and the Variational Quantum Eigensolver are being actively developed and demonstrated in small systems.
However, extremely limited qubit count and low fidelity seriously limit useful applications, especially in the crystalline phase, where compact orbital bases are difficult to develop.
To address this difficulty, we present a hybrid quantum-classical algorithm to solve the band structure of any periodic system described by an adequate tight-binding model.
We showcase our algorithm by computing the band structure of a simple-cubic crystal with one {\em s} and three {\em p} orbitals per site (a simple model for Polonium) using simulators with increasingly realistic levels of noise and culminating with calculations on IBM quantum computers.
Our results show that the algorithm is reliable in a low-noise device, functional with low precision on present-day noisy quantum computers, and displays a complexity that scales as \bigw{M^3} with the number $M$ of tight-binding orbitals per unit-cell, similarly to its classical counterparts.
Our simulations offer a new insight into the ``quantum'' mindset and demonstrate how the algorithms under active development today can be optimized in special cases, such as band structure calculations.
\end{abstract}

\maketitle

\section{Introduction}

The primary objective of computational chemistry is to calculate the eigenstates and eigenenergies of a chemical system.
Increasing the accuracy of these calculations enables more accurate characterization of a wide range of chemical properties (ionization potential, equilibrium constants, IR spectra, etc.).
However, the computational complexity of the most accurate algorithms scales exponentially with the number of basis orbitals, when using a classical computer.
Therefore, large systems are typically treated approximately with a single-electron approximation, in which each electron independently interacts with an effective potential produced by all other electrons and atomic centers.
Spin exchange forces can be treated through the Hartree-Fock self-consistent field method, and multi-body correlations can be treated through \textit{ad hoc} correction terms as in density functional theory, but these methods fail when applied to highly-correlated systems.
Therefore, computational chemists have begun to turn to {\em quantum} computers to reach higher levels of accuracy.

Quantum computers surpass Hartree-Fock by imposing fermionic statistics onto qubits and including electron correlation terms directly in the system Hamiltonian.
Quantum Phase Estimation (QPE) is a quantum algorithm which extracts eigenstate energies from a simulated system on a noise-resilient quantum computer at a low cost of classical computational resources.\cite{Aspuru_Guzik_2005, Dob_ek_2007}
However, quantum circuits designed for the QPE algorithm tend to require longer coherence times than are available in the era of Noisy Intermediate-Scale Quantum (NISQ) devices, so recent efforts focus on hybrid algorithms which balance quantum and classical resource costs.
In particular, a more popular algorithm for molecular ground-state energy calculations is the Variational Quantum Eigensolver (VQE).\cite{Peruzzo_2014,McClean_2016,Kandala_2017,Cerezo_2020DEC}
Many variants of VQE have arisen in recent literature, including methods such as Variational Quantum Deflation (VQD) capable of exploring excited states.\cite{Higgott_2019}
However, resource costs for interesting systems still tend to exceed those currently available, and there is still some doubt whether error in hybrid algorithms can be made to converge sub-exponentially.\cite{Cerezo_2021,Bittel_2021}

Periodic systems, such as the crystalline phase of a molecule or protein, are an especially interesting arena in the development of quantum algorithms.
On the one hand, translational symmetry over the entire crystal seems to offer extremely powerful ways to reduce the computational complexity. \cite{Manrique_2021}
On the other, periodic systems are typically considered infinite in extent and thus appear to require an exceedingly {\em large} number of resources to adequately approximate.
For example, one may simulate a periodic system of $N$ unit cells, each consisting of $M$ orbitals (see for example Cade et al.\cite{Cade_2020}), where $M$ is typically comparable to the number of orbitals considered in a single molecular simulation, but $N$ is large enough to approximate infinity.
Alternatively, one may adopt a plane-wave basis, for which a quantum circuit is available to efficiently diagonalize the kinetic and potential operators directly.\cite{Babbush_2018}
In either approach, the size of the basis, and therefore the number of qubits, must be very large to accurately represent the periodic system.
The quantum resources required to simulate a crystal will thus tend to be many times larger than those required for a solitary molecule, and generally larger than the size of quantum computers available today.

In this work, we offer an easier transition to adapt quantum algorithms for periodic systems, by implementing correlation-free band structure calculations on NISQ-era quantum computers.
Band structures are the fundamental toolbox of materials scientists in the characterization and discovery of the electronic properties of crystalline solids.
Band theory adopts the single-electron approximation (as in Hartree-Fock), for which a periodic Hamiltonian becomes separable in reciprocal space, reducing the system at any particular momentum $\vb{k}$ to the complexity of a single unit cell.
In this way, the eigenstates of an electron with momentum $\vb{k}$ can be efficiently calculated in a classical computer; the energies of each eigenstate along a path varying $\vb{k}$ through reciprocal space form the band structure of the crystal.
Integrating the band structure provides early insight into structural, electronic, optical, and thermal properties of the crystal.\cite{Grosso_2000}

Quantum algorithms to obtain classically verifiable results provide an invaluable tool in establishing foundational building blocks such as efficient mesaurement protocols and error mitigation, \cite{Google_2020AUG} as well as being generally easier to understand and replicated by researchers just breaking into quantum computation.
With this motivation, we show how the single-electron approximation accommodates a systematic approach to efficiently apply VQE to solve the band structure of any periodic system.
We will illustrate our procedure using an empirical tight-binding model for a simple cubic lattice, but our procedure is easily applied to any tight-binding Hamiltonian.
For example, recent work in our group has demonstrated that ``exact'' tight-binding Hamiltonians can be derived from accurate electronic structure calculations using a procedure based on the projection of electronic wavefunctions on localized orbital bases (PAOFLOW).\cite{BuongiornoNardelli_2018}
PAOFLOW is an advanced software tool that constructs tight-binding Hamiltonians from self-consistent electronic wavefunctions projected onto a set of atomic orbitals and provides numerous materials properties that otherwise would have to be calculated via approximate model approaches.
Thus, an efficient quantum algorithm for the solution of the tight-binding Hamiltonian would provide an avenue for the calculation of advanced crystalline properties on a quantum computer.
We do not expect our approach in this paper to offer immediate quantum advantage; rather, our purpose is to help chemists think the {\em quantum} way, motivating new, resource-efficient approaches to studying highly correlated systems.
By considering the simplest available model, we can provide lower bounds on resource complexity and give insight into the practical difficulties chemists may expect when implementing quantum algorithms.

We have previously considered this topic, employing a VQE-based algorithm to iteratively calculate the band energies of a tight-binding silicon model.\cite{Cerasoli_2020}
We now apply recent developments in the literature\cite{Higgott_2019,Gard_2020,Giurgica_Tiron_2020} to extend and improve upon our previous work.
In Section~\ref{sec:Theory}, we briefly outline the essential ideas we have taken from the literature.
In Section~\ref{sec:Method}, we present our robust procedure for accurately calculating the band structure of any periodic system with a quantum computer.
In Section~\ref{sec:Results}, we demonstrate our procedure applied to a simple-cubic lattice, presenting data from a quantum simulator and preliminary results from IBM's open-access \verb|ibmq_athens| and \verb|ibmq_santiago| cloud devices.
In Section~\ref{sec:Discussion}, we discuss the algorithmic complexity of our procedure and highlight the steps which may or may not be improved in later work.

\section{Background}
\label{sec:Theory}
In this section we briefly outline some essential techniques actively studied in the quantum computing literature.
In particular, while QPE provides a robust strategy for measuring eigenenergies with minimal classical resources, its performance suffers greatly from the imperfect fidelity of NISQ devices.
As such, we will focus mostly on VQE and its close cousin VQD when measuring eigenspectra, applying QPE when available as an optional refinement.

\subsection{Quantum Phase Estimation}
\label{sec:Theory:QPE}
In the QPE algorithm,\cite{Aspuru_Guzik_2005, Dob_ek_2007} a set of qubits (the ``state register'') are first prepared into an eigenstate $\ket{\psi}$ of a unitary operator $\hat{U}$, such that $\hat{U}\ket{\psi}=e^{2\pi i \phi}\ket{\psi}$.
The unitary operator $U$ is then repeatedly applied as a controlled quantum circuit so that the phase shift $\phi$ is encoded into another set of qubits (the ``readout register'').
Measurements on the readout register give the binary expansion of $\phi$.
In molecular simulations one selects $\hat{U}\equiv\exp(i\hat{H}\tau)$, the operator which evolves a system with Hamiltonian $\hat{H}$ by a unit time $\tau$.
If one first transforms $\hat{H}$ to guarantee that all possible energies $E$ fall within the interval $[0,2\pi/\tau)$, the measured phases $\phi$ map directly onto the eigenenergies of $\hat{H}$.
The algorithm is deterministic when the state register is prepared exactly into an eigenstate $\ket{\psi}$ and its eigenvalue $\phi$ has an exact binary expansion, but it retains some probability of success when both conditions are relaxed.
Thus, QPE can be adapted to discover eigenstates and eigenvalues \textit{a priori}, at the cost of additional rounds of measurement.

Generally, $\hat{H}$ is given as a weighted sum of non-commuting Pauli words (Section~\ref{sec:Method:H}).
An exact circuit for $\hat{U}\equiv\exp(i\hat{H}\tau)$ is not readily available, but it can be closely approximated by Suzuki-Trotter expansion, which factors $\hat U$ into many small-time slices.\cite{Hatano_2005}
The number of time slices scales polynomially with the accuracy required, and the depth of each time slice depends on the number of commuting groups in $\hat{H}$.
For this reason, QPE is extremely susceptible to errors arising from the low gate fidelity and short coherence times which plague NISQ devices.
Alternative approaches scale more favorably with error at the cost of additional ancilla qubits.\cite{Berry_2015,Kalev_2020}

\subsection{Variational Quantum Eigensolver}
\label{sec:Theory:VQE}
In the VQE algorithm,\cite{Peruzzo_2014, McClean_2016} one begins with a Hamiltonian $\hat{H}$, represented as a weighted sum of non-commuting Pauli words, and a parameterized quantum circuit $\hat{V}(\vb*{\theta})$, the ``ansatz'', which prepares a set of qubits into an arbitrary state (Section~\ref{sec:Method:V}).
The ansatz is applied on an ensemble of states such that qubit measurements give the expectation value of each Pauli word in $H$.
The weighted sum of expectation values gives the energy $E(\vb*{\theta})$ of the arbitrary state prepared - this procedure is called ``operator estimation''.\cite{McClean_2016,Rubin_2018}
The parameters $\vb*{\theta}$ are varied by a classical optimization routine until $E$ is minimized.
According to the variational principle, this minimum is exactly the ground-state energy of the system when the ansatz $\hat{V}(\vb*{\theta})$ is robust (ie. it spans the full Hilbert space of the system), and if the classical optimization succeeds in producing the global minimum.

The algorithmic complexity of VQE depends on several factors.
Measuring the expectation value of a Pauli word is a stochastic process, requiring a large number of measurements on the order of \bigo{\epsilon^{-2}} for an acceptable sampling noise $\epsilon$.
These ensembles are usually measured for each Pauli word in $\hat{H}$, although recent advances reduce the size of the ensemble by simultaneously measuring each commuting group of Pauli words\cite{Yen_2020} or by ``classically shadowing''\cite{Huang_2020} the quantum circuit to require only a logarithmic number of measurements.
Like in QPE, the efficiency of the algorithm is determined by the complexity of $\hat{H}$, and every element of the ensemble requires a unique application of the ansatz, meaning that circuit depth and gate count should be kept minimal.
The dimension of the ansatz also determines the efficiency and efficacy of the classical optimization.
For all these reasons, VQE tends to be impractical for perfectly robust ansatze, and much of the literature focuses on methods for constructing effective ansatze accounting for system symmetries and hardware limitations.\cite{Romero_2018,Ryabinkin_2018,Grimsley_2019,Yen_2020,Gard_2020}
Because circuit depth and gate count are kept low, VQE is well-suited to NISQ devices.

\subsection{Variational Quantum Deflation}
\label{sec:Theory:VQD}
Variational Quantum Deflation (VQD) \cite{Higgott_2019} is one approach for extending the VQE algorithm to explore excited states in addition to the ground-state.
VQD begins as a typical VQE run to locate the ground state, and the ground-state parametrization $\vb*{\theta_0}$ is recorded.
The variational process is then repeated with an additional term in the optimization routine's cost function, which gives the overlap between the current ansatz and the ground-state, weighted by a factor $\beta$.
States similar to the ground-state will be shifted into a higher effective energy, so that the optimization routine considers them unfavorable.
Meanwhile, higher-energy eigenstates must be orthogonal to the ground-state, so their overlap contribution will be zero.
Therefore, the next lowest energy that can be found is the first excited state.
This process is repeated for each energy level, adding a new overlap term for each eigenenergy already found.
Each overlap can be evaluated as the expectation value of a single commuting group of Pauli words in the Hamiltonian, so that the total number of additional measurements after finding $M$ eigenvalues is \bigq{M^2}.
If $\hat{H}$ consists of \bigw{M} commuting groups, measured for each of $M$ energy levels, then the additional cost of the overlap circuits is negligible.

\section{Method}
\label{sec:Method}
Our objective is to calculate the band structure of a periodic system, as described by a tight-binding Hamiltonian $\hat H$ of the form:
\begin{equation}
\label{eq:H:TB}
    \hat H = \sum_{\alpha,\beta} t_{\alpha\beta} c^\dagger_\alpha c_\beta
\end{equation}
Each $c^\dagger_j$ and $c_j$ represent a creation and annihilation (ladder) operator on an atomic orbital $\phi_j$, centered on a coordinate $\vb{r}_j$ in the crystal.
The hopping parameters $t_{\alpha\beta}$ denote the energy cost of an electron transitioning from orbital $\phi_\beta$ to orbital $\phi_\alpha$.
They are calculated from the overlap integrals between each pair of orbitals $\phi_\alpha$ and $\phi_\beta$, or they are selected to fit empirical observations.
A general tight-binding Hamiltonian may also include multi-electron correlation such as $t_{\alpha\beta\gamma\delta} c^\dagger_\alpha c^\dagger_\beta c_\gamma c_\delta$, but we neglect these terms in this work.

Our strategy is to transform Eq.~\ref{eq:H:TB} into reciprocal space and to apply VQD to solve for each eigenenergy at each momentum $\vb{k}$ along the desired path through reciprocal space.
When sufficient quantum resources are available, we refine each band energy with QPE.
Our procedure for mapping a single-electron periodic system onto a set of qubits is derived in Section~\ref{sec:Method:H}.
The variational ansatz we have selected, suitable for any band structure calculation, is described in Section~\ref{sec:Method:V}.
Details of implementing the quantum algorithm are presented in Section~\ref{sec:Method:Algorithm}.
Finally, we provide a step-by-step schematic of our algorithm and its relation to VQE in Fig.~\ref{fig:algorithm}.

\subsection{Qubit Mapping}
\label{sec:Method:H}
The Hamiltonian in Eq.~\ref{eq:H:TB} consists of ladder operators acting on atomic orbitals.
The Hamiltonians appearing in the quantum algorithms of Section~\ref{sec:Theory} consist of Pauli words acting on qubits.
We define a ``Pauli word'' $\hat P_i$ as an operator acting independently on each qubit with either the identity $\hat I$ or one of the Pauli spin matrices $\hat X$, $\hat Y$, $\hat Z$.
Pauli words are a natural choice for representing physical operators in a quantum computer because their expectation values can be readily measured\cite{Cerasoli_2020} and their unitary time evolution $\exp(i\hat P_i t)$ can be readily implemented as a quantum circuit.\cite{Seeley_2012}
Our goal in this section is to map our atomic orbitals onto a qubit basis, and our Hamiltonian to a weighted sum of Pauli words:
\begin{equation}
    \hat H = \sum_\alpha \sum_\beta t_{\alpha\beta} c^\dagger_\alpha c_\beta  \rightarrow \sum_i a_i \hat P_i
\end{equation}

\subsubsection{Qubit Basis}
The simplest conceivable mapping between orbitals and qubits is to identify each orbital with its own qubit. The qubit state $\ket{1}$ represents an occupied orbital, while $\ket{0}$ is empty.
There are however an infinite number of orbitals in an infinite crystal, and quantum computers with an infinite number of qubits are beyond our engineering capabilities. % Kyle still likes 'present-day engineering capabilities', but Frank doesn't.
We therefore reinterpret $\hat H$ as the Hamiltonian of an arbitrarily large supercell with periodic boundary conditions, consisting of $N$ unit cells, each with $M$ orbitals.
\begin{equation}
\label{eq:H:xspace}
    \hat H = \sum_{\vb*{\nu}}^{(N)} \sum_{\vb*{\nu}'}^{(N)} \sum_{\alpha=0}^{M-1} \sum_{\beta=0}^{M-1}  t_{\alpha\beta}^{(\vb*{\delta})} c^\dagger_{\vb*{\nu}'\alpha} c_{\vb*{\nu}\beta}
\end{equation}
Hopping parameters are now dependent on the orbitals $\alpha$, $\beta$ and the displacement vector $\vb*{\delta} \equiv \vb{r}_{\vb*{\nu}'\alpha} - \vb{r}_{\vb*{\nu}\beta}$ between their atoms.
As $\vb*{\delta}$ increases, $t_{\alpha\beta}^{(\vb*{\delta})}$ tends to vanish, permitting a nearest-neighbor approximation in which one considers only a few of the smallest $\vb*{\delta}$.

Eq.~\ref{eq:H:xspace}, when supplemented with two-electron correlation terms, is the form typically considered when applying quantum algorithms to periodic systems, requiring a total of $MN$ qubits.
In the single-electron approximation, however, we can reduce the size of the system to only $M$ qubits by transforming into reciprocal space.
Reciprocal space orbitals are characterized by their own ladder operators $\tilde c^\dagger_{\vb{k}j}$ and $\tilde c_{\vb{k}j}$, related to $c^\dagger_{\vb*{\nu}j}$ and $c_{\vb*{\nu}j}$ by Fourier transform:
\begin{subequations}
\label{eq:c:fourier}
\begin{align}
    c^\dagger_{\vb*{\nu}'\alpha} &= \frac{1}{\sqrt{N}} \sum_{\vb{k}'} e^{ i\vb{k}'\vdot\vb{r}_{\vb*{\nu}'\alpha}} \tilde c^\dagger_{\vb{k}'\alpha} \\
    c        _{\vb*{\nu}\beta} &= \frac{1}{\sqrt{N}} \sum_{\vb{k}} e^{-i\vb{k}\vdot\vb{r}_{\vb*{\nu}\beta}} \tilde c        _{\vb{k}\beta}
\end{align}
\end{subequations}
Substituting Eqs.~\ref{eq:c:fourier} into Eq.~\ref{eq:H:xspace}, we obtain
\begin{align}
    \hat H = \sum_{\vb{k}}^{(N)} & \sum_{\vb{k}'}^{(N)} \sum_{\alpha=0}^{M-1} \sum_{\beta=0}^{M-1} \qty( \frac{1}{N} \sum_{\vb*{\nu}} \sum_{\vb*{\nu}'} t_{\alpha\beta}^{(\vb*{\delta})} e^{i(\vb{k}'\vdot\vb{r}_{\vb*{\nu}'\alpha} - \vb{k}\vdot\vb{r}_{\vb*{\nu}\beta})} \tilde c^\dagger_{\vb{k}'\alpha} \tilde c_{\vb{k}\beta} )
\end{align}
We simplify this sum by recalling $\vb{r}_{\vb*{\nu}'\alpha} = \vb{r}_{\vb*{\nu}\beta} + \vb*{\delta}$.
Then $\vb{r}_{\vb*{\nu}'\alpha}$ becomes a common factor of each $\vb{k}$ in the exponential, and we may exploit the orthogonality relation $\frac{1}{N} \sum_{\vb*{\nu}'} e^{i(\vb{k}' - \vb{k}) \vdot \vb{r}_{\vb*{\nu}'\alpha}} = \delta_{\vb{k}'\vb{k}}$.
Summing over $\delta_{\vb{k}'\vb{k}}$, we obtain $\hat H = \sum_{\vb{k}} \hat H_{\vb{k}}$, where
\begin{align}
\label{eq:H:kspace}
\hat H_{\vb{k}} &\equiv \sum_{\alpha=0}^{M-1} \sum_{\beta=0}^{M-1} H_{\alpha\beta}(\vb{k}) \tilde c^\dagger_{\vb{k}\alpha} \tilde c_{\vb{k}\beta} \\
\label{eq:H:matrix}
    H_{\alpha\beta}(\vb{k}) &\equiv \sum_{\vb*{\delta}} t_{\alpha\beta}^{(\vb*{\delta})} e^{i \vb{k} \vdot \vb*{\delta} }
\end{align}
Each momentum $\vb{k}$ contributes an independent subystem with only $M$ orbitals, whose eigenenergies may be solved independently.
Classically, the values $H_{\alpha\beta}(\vb{k})$ in Eq.~\ref{eq:H:matrix} form an $M\times M$ Hermitian matrix whose eigenvalues can be efficiently calculated with standard linear algebraic techniques in \bigq{M^3} time.
This work instead considers how to calculate these eigenvalues the ``quantum'' way.

We focus on a specific $\hat H_{\vb{k}}$ for the remainder of this section, with the understanding that our procedure must be repeated for each momentum $\vb{k}$ along the path of interest in reciprocal space.
Eq.~\ref{eq:H:kspace} has a form very similar to Eq.~\ref{eq:H:TB}, except that it acts on reciprocal-space orbitals rather than atomic orbitals.
We therefore adopt a ``reciprocal-orbital'' basis, in which each reciprocal-space orbital is identified with its own qubit.

\subsubsection{Hamiltonian Mapping}
Having transformed our Hamiltonian into reciprocal space (Eq.~\ref{eq:H:kspace}), we must now consider mapping each ladder operator to a set of Pauli words.
The ladder operators must satisfy the following:
\begin{subequations}
\begin{align}
    \tilde c         \ket{0} &=  0       &
    \tilde c         \ket{1} &=  \ket{0} \\
    \tilde c^\dagger \ket{0} &=  \ket{1} &
    \tilde c^\dagger \ket{1} &=  0 &
\end{align}
\end{subequations}
Meanwhile, the Pauli spin operators $\hat X$, $\hat Y$, $\hat Z$ act on a qubit's basis states in the following way:
\begin{subequations}
\begin{align}
    \hat X \ket{0} &= \ket{1} &
    \hat X \ket{1} &= \ket{0} \\
  -i\hat Y \ket{0} &= \ket{1} &
   i\hat Y \ket{1} &= \ket{0} \\
    \hat Z \ket{0} &= \ket{0} &
  - \hat Z \ket{1} &= \ket{1}
\end{align}
\end{subequations}
It is easy to verify that the following mapping suffices for a single qubit:
\begin{subequations}
\label{eq:mapping:c}
\begin{align}
    \tilde c         &\rightarrow \frac{1}{2}( \hat X + i \hat Y ) \\
    \tilde c^\dagger &\rightarrow \frac{1}{2}( \hat X - i \hat Y )
\end{align}
\end{subequations}

In multi-electron systems, one typically adopts the Jordan-Wigner transformation, which retains the form of Eqs.~\ref{eq:mapping:c} but appends a $\hat Z$ operation on \bigq{M} other qubits to enforce fermionic antisymmetry.
Alternatively, one may adopt the Bravyi-Kitaev transformation, which requires operations on only \bigq{\log M} qubits, but uses a non-intuitive basis and involves non-adjacent interactions more difficult to simulate on certain qubit architectures.
We refer the reader to Seeley et al.\cite{Seeley_2012} for an excellent introduction to both transforms.
However, because we are considering single-electron systems, there {\em are} no other fermions to exchange with, and we may use Eqs.~\ref{eq:mapping:c} directly, so that each ladder operator acts on only \bigq{1} qubits.
This significantly reduces the complexity of our Hamiltonian, as we shall presently see.

We may rewrite Eq.~\ref{eq:H:kspace} to exploit the Hermiticity of $\hat H_{\vb{k}}$.
\begin{equation}
    \label{eq:H:kspace:adjoint}
    \hat H_{\vb{k}} = \sum_\alpha H_{\alpha\alpha} c^\dagger_\alpha c_\alpha + 2 \sum_\alpha \sum_{\beta>\alpha} \Re(H_{\alpha\beta} c^\dagger_\alpha c_\beta)
\end{equation}
since the transpose term $ H_{\beta\alpha} c^\dagger_\beta c_\alpha = (H_{\alpha\beta} c^\dagger_\alpha c_\beta)^\dagger $.
Applying Eqs.~\ref{eq:mapping:c} and noting $\hat X^2=\hat Y^2=\hat I$, $-i\hat X\hat Y = i\hat Y\hat X = \hat Z$:
\begin{align}
\label{eq:H:Pauli}
    \hat H_{\vb{k}} \rightarrow& \frac{1}{2} \sum_\alpha H_{\alpha\alpha} (\hat I - \hat Z_\alpha) \nonumber \\
    &+ \frac{1}{2} \sum_\alpha \sum_{\beta>\alpha} \Re(H_{\alpha\beta}) (\hat X_\alpha \hat X_\beta + \hat Y_\alpha \hat Y_\beta) \nonumber \\
    &+ \frac{1}{2} \sum_\alpha \sum_{\beta>\alpha} \Im(H_{\alpha\beta}) (\hat Y_\alpha \hat X_\beta - \hat X_\alpha \hat Y_\beta)
\end{align}
Eq.~\ref{eq:H:Pauli} provides the weighted sum of Pauli words required in the quantum algorithms of Section~\ref{sec:Theory}.

Eq.~\ref{eq:H:Pauli} consists of \bigq{M^2} Pauli words.
The complexity of each algorithm in Section~\ref{sec:Theory} is determined in part by the number of commuting groups in $\hat H$.
In Eq.~\ref{eq:H:Pauli}, all terms of the form $\hat Z_\alpha$, $\hat X_\alpha \hat X_\beta$, and $\hat Y_\alpha \hat Y_\beta$ each form commutative groups.
Therefore, when $\hat H_{\vb{k}}$ has no imaginary part, the energy can be determined with just 3 rounds of measurement.
When $\hat H_{\vb{k}}$ does have an imaginary part, we note that for fixed $\alpha$, $\hat Y_\alpha \hat X_{\beta>\alpha}$ and $\hat X_\alpha \hat Y_{\beta>\alpha}$ each form commutative groups, so in general we have \bigq{M} commuting groups.
Finally, we note that each of these commuting groups are {\em qubit-wise} commutative, meaning that each index of all Pauli words in the set has either the same spin operator or the identity.
This accommodates a particularly simple procedure for measuring expectation values of each set simultaneously, requiring no additional overhead in the measurement circuit.

\begin{figure*}[ht] \centering
\begin{subfigure}{\textwidth} \centering
\begin{quantikz}
    \lstick{$\ket{0}$}
        & \qw
        & \gate{X}
            \gategroup[wires=4,steps=7,background,style={
                rounded corners,
                % fill=gray!20
            }]{}
        & \qw
        & \gate[wires=2]{A(\theta_1,\phi_1)}
        & \qw
        & \qw
        & \qw
        & \qw
        & \qw \\
    \lstick{$\ket{0}$}
        & \qw
        & \qw
        & \qw
        & \qw
        & \qw
        & \gate[wires=2]{A(\theta_2,\phi_2)}
        & \qw
        & \qw
        & \qw \\
    \lstick{$\ket{0}$}
        & \qw
        & \qw
        & \qw
        & \qw
        & \qw
        &
        & \ \ldots\ \qw
        & \gate[wires=2]{A(\theta_{M-1},\phi_{M-1})}
        & \qw \\
    \lstick{$\ket{0}$}
        & \qw
        & \qw
        & \qw
        & \qw
        & \qw
        & \qw
        & \qw
        &
        & \qw
\end{quantikz}
\caption{\centering The ansatz $\hat V(\vb*{\theta})$.}
\end{subfigure}
\begin{subfigure}{\textwidth} \centering
\begin{quantikz}
        & \ctrl{1}
            \gategroup[wires=2,steps=7,background,style={
                % fill=gray!20
            }]{}
        & \gate{R_z^\dagger(\phi+\pi)}
        & \gate{R_y^\dagger(\theta+\frac{\pi}{2})}
        & \targ{}
        & \gate{R_y(\theta+\frac{\pi}{2})}
        & \gate{R_z(\phi+\pi)}
        & \ctrl{1}
        & \qw \\
    % next line
        & \targ{}
        & \qw
        & \qw
        & \ctrl{-1}
        & \qw
        & \qw
        & \targ{}
        & \qw
\end{quantikz}
\caption{\centering The particle-number preserving $A(\theta,\phi)$ gate from Gard et al.\cite{Gard_2020}}
\end{subfigure}

    \caption{\label{fig:ansatz}The ansatz $\hat V(\vb*{\theta})$ suitable for any band structure calculation. Each qubit is initialized in the $\ket{0}$ state; the output is an arbitrary superposition of states with a single qubit in the $\ket{1}$ state.}
\end{figure*}
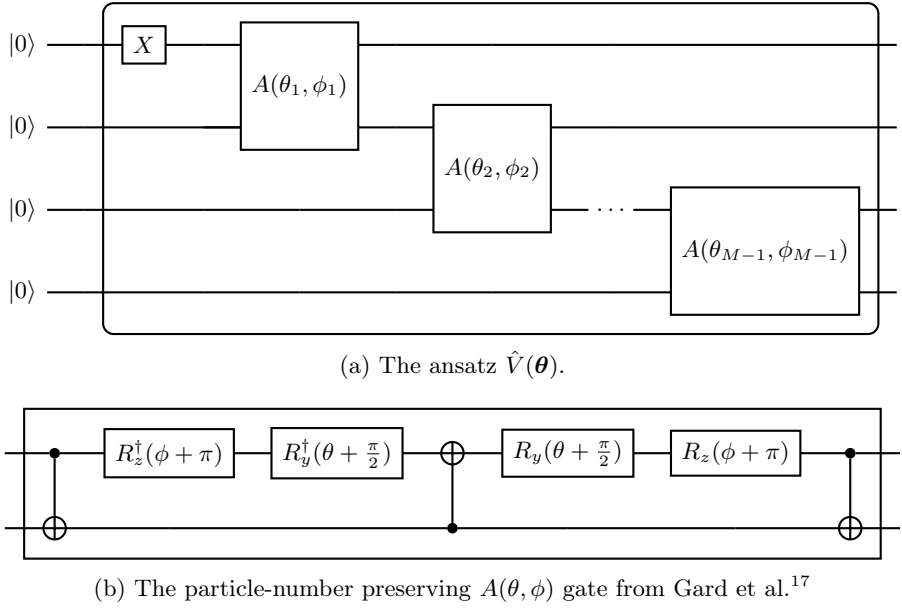

\subsection{Ansatz}
\label{sec:Method:V}
The VQE and VQD algorithms require an ansatz - a parameterized quantum circuit $\hat V(\vb*{\theta})$ preparing a trial state $\Psi(\vb*{\theta})\equiv \hat V(\vb*{\theta})\ket{0}$ for energy measurements.
A quantum circuit to span the full Hilbert space of $M$ qubits requires $2(2^M-1)$ parameters, and it will not generally have an efficient decomposition into one- and two-qubit gates.
However, most applications to molecular simulation consider a system with fixed number of electrons.
In the orbital basis, or in our reciprocal orbital basis, one need only consider that subset of Hilbert space spanned by the basis states whose Hamming weights match the number of electrons in our system.
For example, in band structure calculations we consider just one electron, so we need only consider the space spanned by $\ket{10...}$, $\ket{010...}$, etc..

Gard et al.\cite{Gard_2020} provide a procedure for generating variational ansatz which conserve particle number, which is particularly simple when particle number is 1.
We begin with $M$ qubits labeled $0$ through $M-1$ in the state $\ket{0}$.
First, we apply an $\hat X$ gate to qubit $0$, to set our ansatz with a single filled orbital.
Then we apply the entangling parameterized $A$ gate\cite{Gard_2020} such that each qubit is entangled directly or indirectly with qubit $0$ (see Fig.~\ref{fig:ansatz}).
This ansatz requires $M-1$ $A$ gates, each contributing two independent parameters, for a total of \bigq{M} gates and parameters.
The circuit is compatible with any quantum architecture exhibiting linear qubit connectivity and has a depth of \bigq{M}.
Alternatively, in a fully connected device, the $A$ gates could be applied with a ``divide-and-conquer'' strategy, reducing the circuit depth to \bigw{\log M}.

Rather than assigning each orbital to its own qubit, we {\em could} assign each orbital to an individual basis state, requiring only \bigq{\log M} qubits total.
This ``compact basis'' is the approach of our previous work.\cite{Cerasoli_2020}
While this is more efficient in the number of qubits, it must explore states with an arbitrary Hamming weight.
The number of parameters required to span the space of interest is unchanged, and a suitable ansatz must be developed \textit{ad hoc}.
Additionally, the Hamiltonian for the compact basis consists of global operators acting on all qubits at once, and it would generally form the maximum number $3^{\log_2 M}=M^{\log_2 3}$ of commuting sets, requiring a less efficient measurement protocol.
Reliance on a random ansatz and a global cost function made our previous procedure vulnerable to exponentially difficult optimizations induced by barren plateaus.\cite{McClean_2018,Cerezo_2021}
Finally, the Hamiltonian for the compact basis is less-structured and more difficult to reduce based on symmetries in the Hamiltonian (for example our observation in Section~\ref{sec:Method:H} that a real $\hat H_{\vb{k}}$ results in \bigq{1} rounds of measurement).

\begin{figure*}[ht] \centering
 \includegraphics[width=\linewidth]{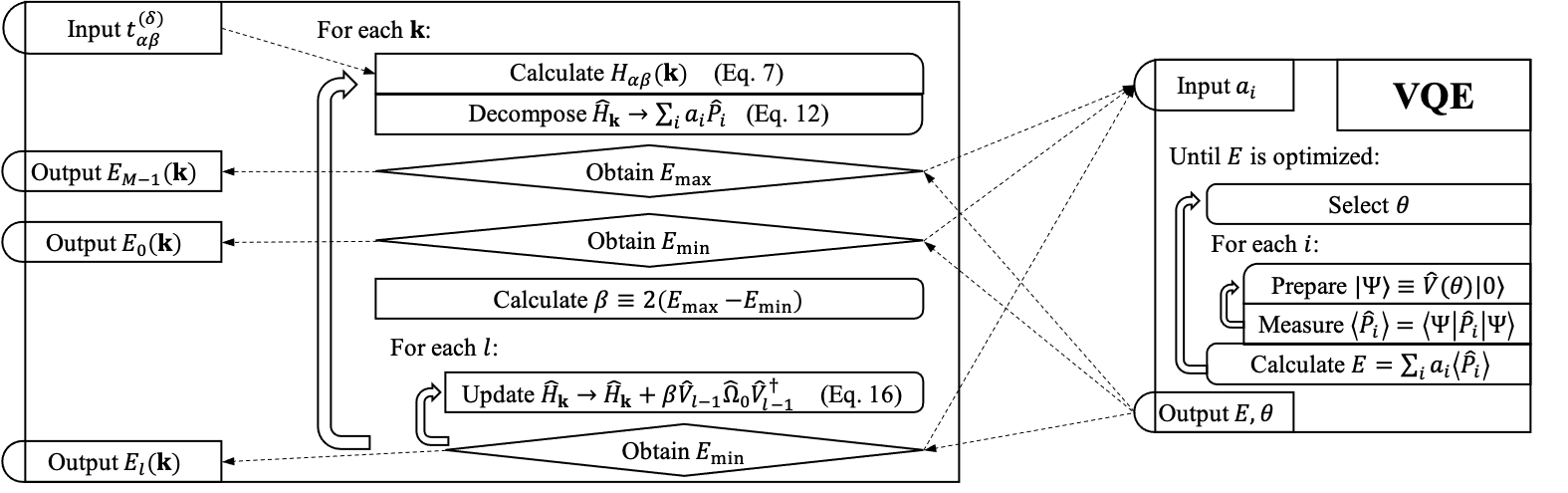}
 \caption{\label{fig:algorithm} A schematic of our algorithm and its relation to VQE. Our algorithm takes tight-binding parameters $t_{\alpha\beta}^{(\vb*{\delta})}$ as input and outputs each band energy $E_l(\vb*{\theta})$. Optionally, each band energy may be refined with QPE. The operator $\hat \Omega_0$ is the sum of all Pauli words spelled with letters $\hat I$ and $\hat Z$. The operator $\hat V$ is the quantum circuit presented in Fig.~\ref{fig:ansatz}.}
\end{figure*}

\subsection{Band Structure Calculation}
\label{sec:Method:Algorithm}

With our ansatz $\hat V(\vb*{\theta})$ (Fig.~\ref{fig:ansatz}) and qubit Hamiltonian $\hat H_{\vb{k}}$ (Eqs.~\ref{eq:H:matrix} and~\ref{eq:H:Pauli}) prepared, we are ready to implement VQD for each momentum $\vb{k}$ along a path through reciprocal space.
This path is usually constructed from high-symmetry segments in the crystal's First Brillouin Zone, because this proves sufficient to calculate many properties of interest.
As briefly described in Section~\ref{sec:Theory}, the idea is to vary the trial state prepared by our ansatz until the energy $E \equiv \expval*{\hat H}$ is minimized.
We repeat the optimization for each band energy, adding additional terms to the cost function proportional to the overlap between the trial state and each previously found eigenstate, weighted by the constant factor $\beta$.

The expectation values $\expval*{\hat H}$ of a generic observable cannot be directly measured in the quantum computer.
Rather, the expectation value of each Pauli word $\hat P_i$ are measured independently, and the energy is evaluated from the weighted sum $\expval*{\hat H} = \sum_i a_i \expval*{\hat P_i}$, with weights $a_i$ taken from Eq.~\ref{eq:H:Pauli}.
Obtaining the Pauli expectation values $\expval*{\hat P_i}$ is also somewhat indirect.
First, the Pauli word $\hat P_i$ should be transformed so that it contains only letters $\hat I$ or $\hat Z$ - let us refer to the modified Pauli word as $\hat Q_i$.
In practice, the transformation is easily accomplished by applying a ``basis rotation'' gate to each qubit before measurement.
Next, each qubit is measured to be in one of the two computational basis states $\ket{0}$ or $\ket{1}$.
The bitstring obtained from concatenating the state of each qubit is itself an eigenstate of $\hat Q_i$, with eigenvalue +1 or~-1.
This procedure is applied to a large {\em ensemble} of qubits, each prepared independently with the ansatz and basis rotation gates.
The expectation value $\expval*{\hat P_i}$ is the average of all the eigenvalues of $\hat Q_i$ measured across the ensemble.

The ensemble necessarily has a finite size $S$, introducing an energy variance on the order of $\epsilon^2 \in$ \bigo{1/S}.
In practice, the ensemble is usually prepared in sequence, resetting a single register of qubits after each round of measurement, relegating the sampling error $\epsilon$ a parameter in the time complexity of any VQE-based algorithm.
Fortunately, the same ensemble may be used to calculate the expectation values of any Pauli word which is {\em qubit-wise} commutative with $\hat P_i$.
For simplicity, we assign $S=8096$ for each commuting group in this work, although advanced methods exist which optimally distribute measurements to minimize the sampling error $\epsilon$.\cite{Rubin_2018}

Many popular optimization routines (eg. SLSQP, BFGS) are gradient-based, and they have difficulty converging to the correct value in the presence of sampling noise.
Therefore, we use COBYLA, a simplex-based algorithm implemented in the \verb|scipy| python package, which we have empirically noted to give good results.
We randomly generate our initial guess for the parameters $\vb*{\theta}$, and we use the default tolerance parameters implemented by \verb|scipy|.
These choices are by far the simplest, but they are by no means optimal, and our results may be improved greatly by a more careful choice of optimization routine.\cite{Lavrijsen_2020}

Before we can implement the deflation procedure, we must select the constant $\beta$ suitable for ``deflating'' each band energy.
We do this with a systematic procedure, first {\em maximizing} the energy of our system to find the highest possible energy $E_{\rm max}$.
We then minimize the energy to find $E_0$ and $\Delta\equiv E_{\rm max}-E_0$.
In theory, $\beta=\Delta$ is a sufficiently high number to guarantee each eigenstate is projected sufficiently out of the optimization in later steps.
In practice, we take $\beta=2\Delta$ to insure against errors in the sampling and optimization process.

Higgott et al.\cite{Higgott_2019} offer several strategies for computing the overlap, offering robustness against error at the cost of ancilla qubits or additional optimization steps.
In this work, we choose the simplest, evaluating the overlap with an eigenstate $\Psi(\vb*{\theta}_l)$ by preparing the trial state $\Psi(\vb*{\theta})$ and applying the {\em adjoint} circuit $\hat V_l^\dagger \equiv \hat V^\dagger(\vb*{\theta}_l)$.
The probability of measuring the bitstring 0..0 gives the overlap $|\braket{\Psi(\vb*{\theta})}{\Psi(\vb*{\theta}_l)}|^2$.
In practice, the probability of measuring bitstring 0..0 is equivalent to the expectation value of an operator $\Omega_0\equiv\sum_i \hat Q_i$, the sum of all unique Pauli words spelled with the letters $\hat I$ and $\hat Z$ (eg. $\hat I \hat I \hat I$, $\hat I \hat I \hat Z$, ... $\hat Z \hat Z \hat Z$).
All such operators are qubit-wise commutative and can be estimated with a single round of measurements.
Therefore, we can implement the deflation procedure conveniently in the \verb|qiskit| Python package provided by IBM, by solving for each band energy and then adding to our Hamiltonian the deflation operator $\beta \hat V_l \hat \Omega_0 \hat V_l^\dagger$.

Initializing the Hamiltonian $\hat H_0 \equiv \hat H_{\vb{k}}$, our procedure can be formally summarized as follows:
\begin{align}
    \vb*{\theta}_l &\equiv \arg\min_{\vb*{\theta}} \expval{\hat V^\dagger(\vb*{\theta}) \hat H_l \hat V(\vb*{\theta})}{0} \\
    \hat V_l &\equiv \hat V(\vb*{\theta}_l) \ket{0} \\
    E_l &\equiv \expval{\hat V_l^\dagger \hat H_l \hat V_l}{0} \\
\label{eq:H:deflation}
    H_{l+1} &\equiv H_l + \beta \hat V_l \hat \Omega_0 \hat V_l^\dagger
\end{align}
Each $E_l$ we find is recorded as the energy of the $l$th band at momentum $\vb{k}$, and we repeat the procedure for each $\vb{k}$ in our selected path.

Optimization routines do not always converge to the true minimum, and errors incurred early in the deflation procedure can propagate unfavorably to higher bands.
Therefore, we include an optional QPE refinement to our algorithm, which applies QPE to each optimized state $\Psi_l \equiv \hat V_l \ket{0}$.
QPE has the effect of selecting the dominant eigenstate of $\Psi_l$ and giving the corresponding eigenenergy with high precision.
Thus, as long as the optimization procedure is ``good enough'', we may update our energy calculations with the result of the QPE experiment.
We have used the iterative version of QPE implemented in \verb|qiskit|.
Details of the algorithm can be found in Dob\u{s}\'{i}\u{c}ek et al.\cite{Dob_ek_2007}

\begin{figure*}[ht] \centering
 \includegraphics[width=\linewidth]{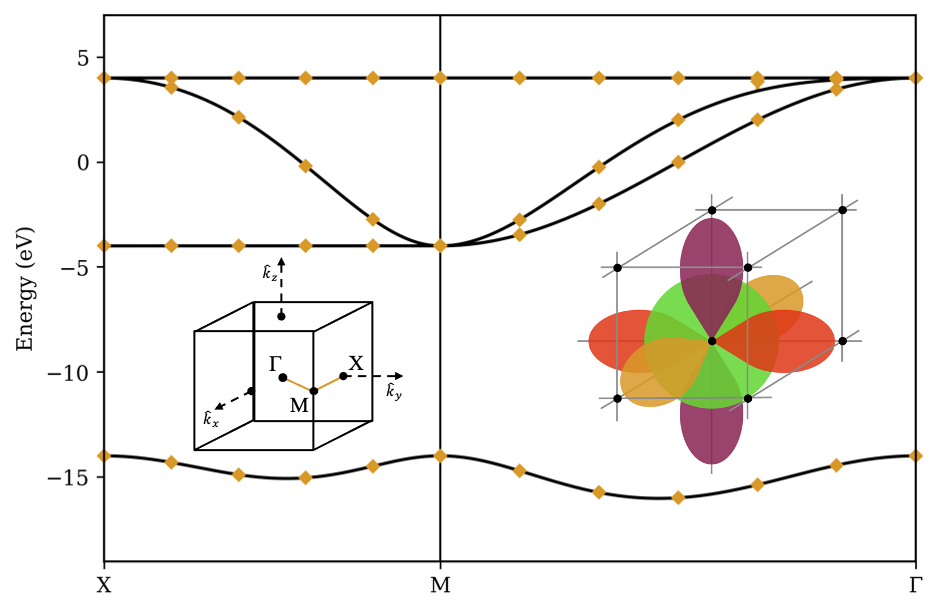}
 \caption{\label{fig:Po:SVM} \textbf{Statevector Simulator} - The band structure of a simple cubic lattice with {\em s} and {\em p} orbitals (right inset) along the high-symmetry path XM$\Gamma$ through the lattice's First Brillouin Zone (left inset). Solid curves denote classical (exact) diagonalization. Diamonds denote the median optimization result from applying our method on a noiseless statevector simulator 32 times with a different random seed. Bars (only visible between the third and fourth bands at nearly-degenerate momenta) denote interquartile ranges. Hopping parameters are 2 eV between adjacent {\em s} and {\em p} orbitals and 2 eV between colinear {\em p} orbitals. Each {\em s} orbital has a self-energy of -14 eV to generate a large band gap.}
\end{figure*}

\section{Results}
\label{sec:Results}

To demonstrate our procedure, we consider a basic model for a crystal in a simple cubic lattic structure (see Fig.~\ref{fig:Po:SVM}).
Each atom has $s$, $p_x$, $p_y$, and $p_z$ orbitals ($M=4$), with a large energy gap between the $s$ and $p$ orbitals, and comparatively small hopping parameters between neighboring $s$ and $p$ orbitals.
This is the simplest three-dimensional periodic system accommodating multiple orbitals per atom, and the required number of qubits (one qubit per orbital, plus one ancilla qubit to implement QPE) fits nicely onto IBM's open-access five-qubit machines.
It may also be considered a rough model for elemental Polonium, although more accurate models should take into account relativistic effects and Coulomb interaction between orbitals located on the same atom.\cite{Min_2006,Silva_2017}
The exact eigenenergies of our model at specific k-points along a high-symmetry path are calculated using standard linear algebraic techniques to diagonalize the matrix elements in Eq.~\ref{eq:H:matrix}.
We compare this band structure to the results from the quantum algorithm presented in Section~\ref{sec:Method} in four different levels of  simulation:
\begin{enumerate}
    \item \textbf{Statevector} - quantum operations are simulated with unitary matrices, and expectation values are calculated exactly.
    \item \textbf{Sampling} - expectation values are now calculated by sampling from a probability distribution.
    \item \textbf{Noisy} - quantum operations and measurements are now applied with an error rate drawn from real quantum devices.
    \item \textbf{Calibrated} - the same noisy simulator is used, but classical post-processing steps are applied to mitigate the error.
\end{enumerate}
We also present preliminary results from IBM quantum devices.

\subsection{Statevector Simulator}

To validate our algorithm's capability of producing the correct band structure, we model the state of an $n$ qubit system as a complex statevector and quantum operations as unitary matrices acting on the Hilbert space spanned by the $2^n$ dimensional basis vectors.
Expectation values are evaluated analytically.
Such a simulation gives the ideal behavior of a quantum computer, with perfect qubit fidelity and no sampling variance.
Fig.~\ref{fig:Po:SVM} summarizes the results of over 32 randomly-seeded optimization runs, marking the median optimization with a diamond and the interquartile range with a bar.
For a few momenta where the third and fourth bands are very close together, optimization tends to locate the wrong eigenstate, giving a small variance in results.
For every other point, the diamond coincides perfectly with the classical solution and the bar is absent, demonstrating that our ansatz is robust, the deflation procedure is mathematically sound, and that our choice of optimization routine (COBYLA) is generally consistent in converging to the correct values on a smooth surface.

\begin{figure*}[ht] \centering
 \includegraphics[width=\linewidth]{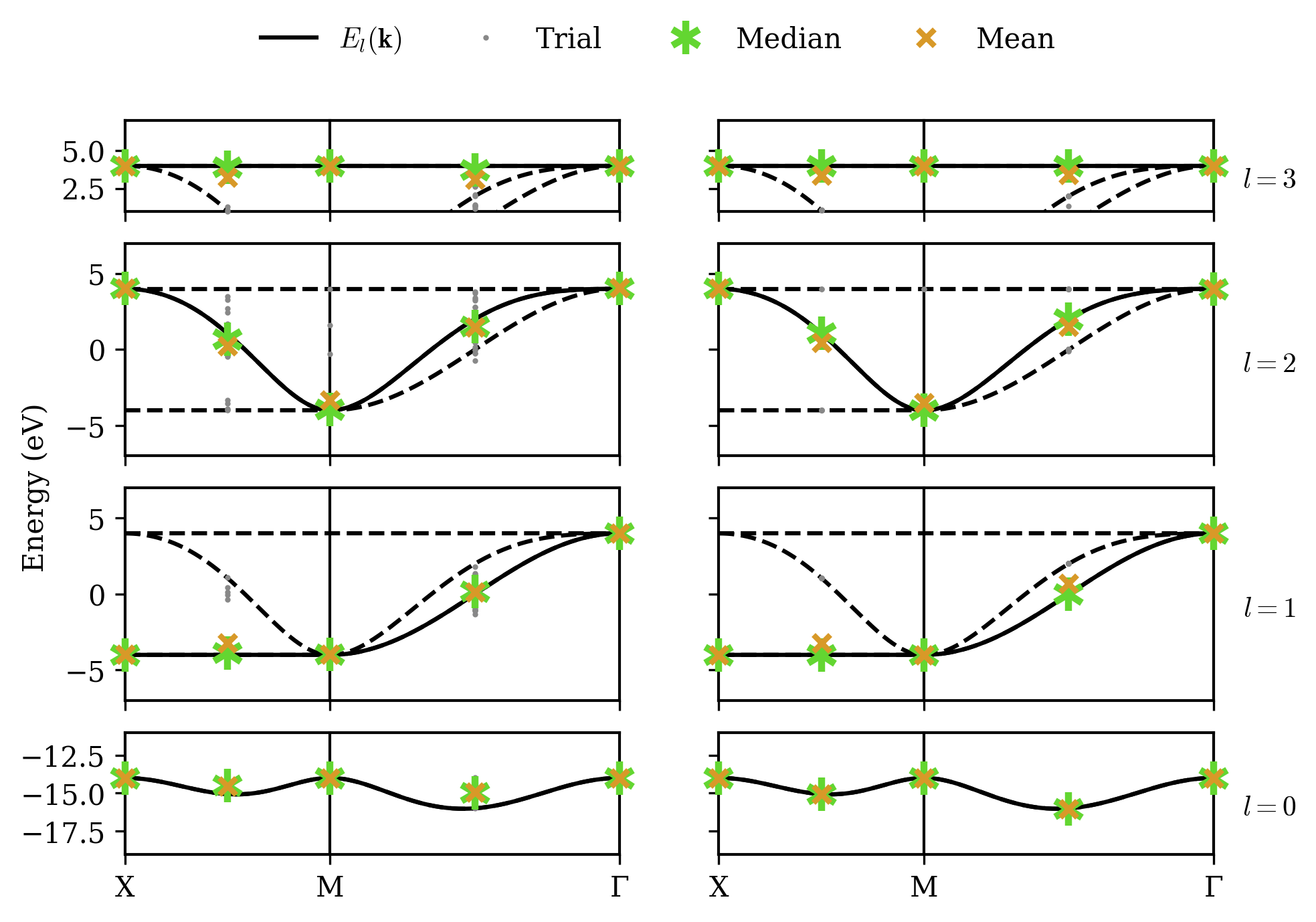}
 \caption{\label{fig:Po:QVM} \textbf{Sampling Simulator} - Our method applied in the presence of sampling noise (high-fidelity qubits). The left column shows raw optimization results; the right column shows the energy obtained by QPE refinement. Gray dots denote the results from each of 32 trials, with each band given its own row. The asterisk and X denote the median and mean, respectively.}
\end{figure*}

\subsection{Sampling Simulator}

We now consider long-term viability of our procedure by retaining perfect qubit fidelity but simulating realistic measurement.
The same unitary matrices as in the statevector simulator are applied to an ensemble of states, which are ``measured'' by sampling from the resulting probability distribution a finite number of times.
While mathematically equivalent, the sampling noise resulting from the stochastic measurement process can make the energy surface bumpy, which can have a detrimental effect on the optimization step.
We have selected the COBYLA optimization algorithm because it is resistant to these bumps; nevertheless, the anomalous variance observed at nearly degenerate points in the statevector simulator is now commonplace.
Fig.~\ref{fig:Po:QVM} shows our results on the noiseless qubit simulator over 32 randomly-seeded runs, clearly marking the median (asterisk) and mean (X) for both optimization (left) and QPE refinement (right).
The smaller dots denote the results of individual trials.

The optimization results are extremely accurate and precise on the high-symmetry momenta but deviate slightly on the intermediate points.
In fact, the high-symmetry points in this particular model each happen to have matrices $H_{\alpha\beta}(\vb{k})$ (Eq.~\ref{eq:H:matrix}) which are already diagonalized, and the resulting cost function yields a well-behaved surface which is reliably optimized, even in the presence of noise.
Averaged results on the intermediate points still tend to be quite good, but individual trials can exhibit a large variance.
However, the optimization does succeed in finding a point {\em close enough} to a correct eigenstate that the QPE refinement consistently extracts the dominant eigenvalue with high precision.
The median QPE results prove to be as accurate as is permitted by the finite binary expansion calculated by the algorithm.

\begin{figure*}[ht] \centering
 \includegraphics[width=\linewidth]{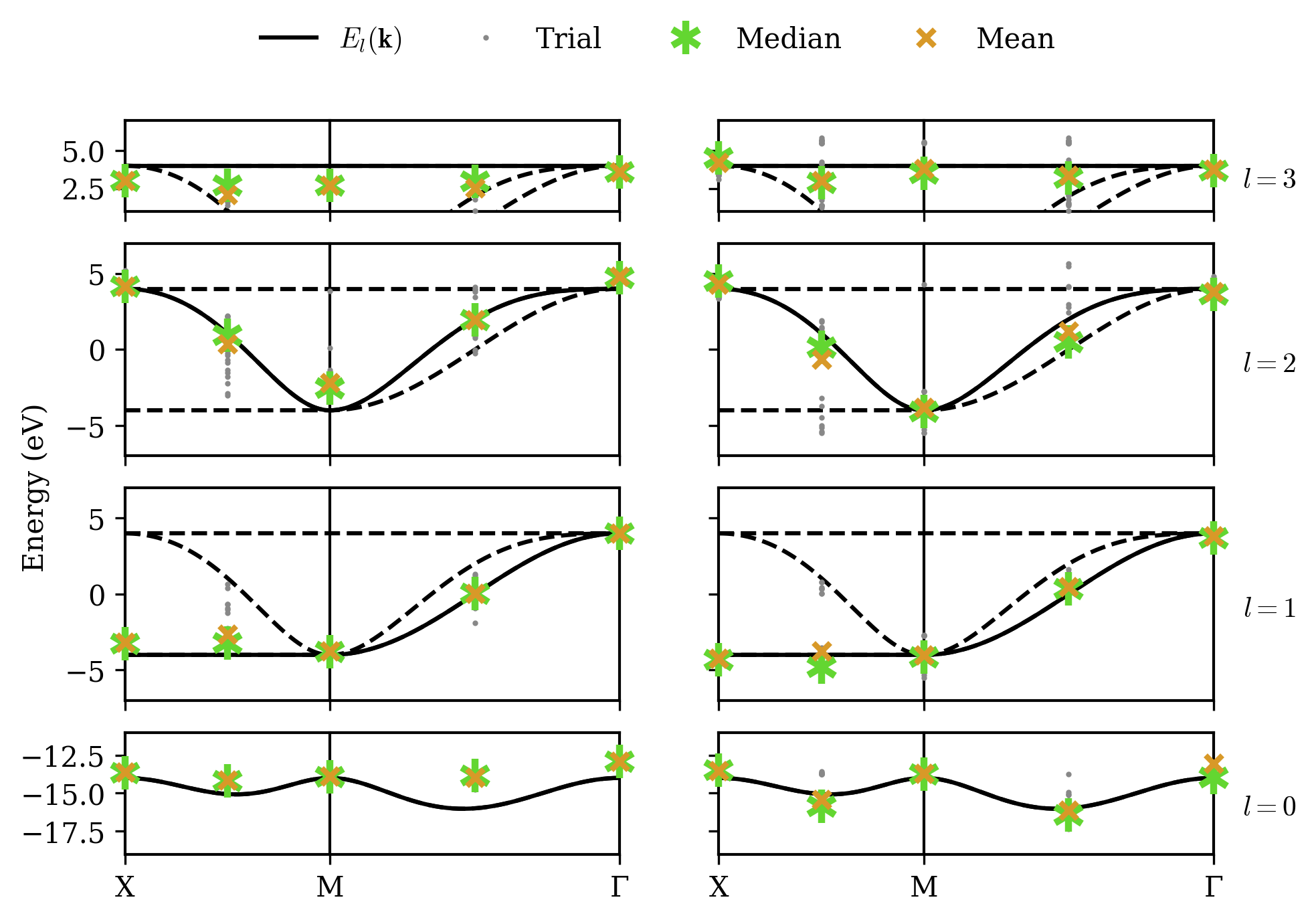}
 \caption{\label{fig:Po:RAW} \textbf{Noisy Simulator} - Our method applied while simulating low-fidelity qubits, without calibration. The left column shows raw optimization results; the right column shows the energy obtained by QPE refinement. Gray dots denote the results from each of 32 trials, with each band given its own row. The asterisk and X denote the median and mean, respectively.}
\end{figure*}

\subsection{Noisy Simulator}

We now consider the realistic application of our procedure on present-day quantum computers, which suffer from relatively short coherence times and are vulnerable to a number of error sources.
This makes practical computations extremely difficult, even in systems requiring relatively few qubits.
We model environmental effects by simulating a dephasing channel, randomly introducing a bit-flip and/or phase-flip on each qubit after each unitary operation \cite{Nielsen_2011}.
We also model errors in the measurement process by randomly introducing a bit-flip with low probability prior to sampling each probability distribution.
Fig.~\ref{fig:Po:RAW} shows our results on a simulator emulating the error rates characteristic of IBM's \verb|ibmq_athens| quantum computer.
Qubit noise has a clearly negative impact on the quality of results.
Lowest-band optimization results tend to suffer a large systematic shift, characteristic of coherent noise in a quantum computer.
Additionally, while average QPE refinement often improves energy estimates, its results are now clustered with some variance around each nearby band, and on occasion (eg. the third band between M and $\Gamma$) the mean optimization result is more accurate.
This is symptomatic of the long circuit requirements for QPE, and supports the widely-believed notion that variational algorithms are better suited to NISQ devices.

\begin{figure*}[ht] \centering
 \includegraphics[width=\linewidth]{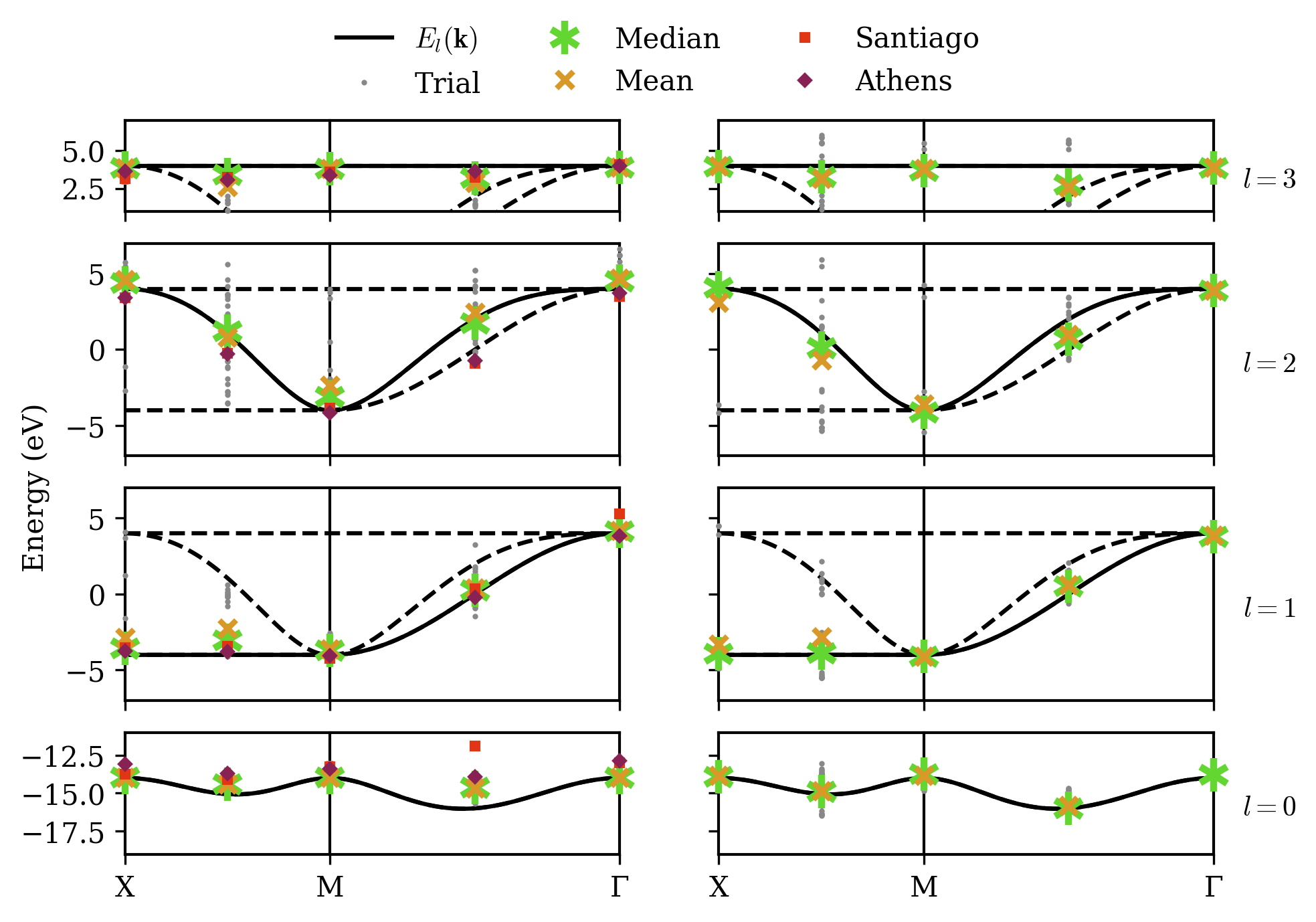}
 \caption{\label{fig:Po:CAL} \textbf{Calibrated Simulator} - Our method applied while simulating low-fidelity qubits, along with rudimentary calibration. The left column shows raw optimization results; the right column shows the energy obtained by QPE refinement. Gray dots denote the results from each of 32 trials, with each band given its own row. The asterisk and X denote the median and mean, respectively. The squares and diamonds on the left denote the energies measured on quantum devices \texttt{ibmq\_santiago} and \texttt{ibmq\_athens} respectively, using the least-error optimization results obtained with the calibrated simulation data. Device architecture constrains the length of quantum circuits, and QPE results for either device could not be obtained.}
\end{figure*}

\subsection{Calibrated Simulator}

Although automated error correction procedures, based on redundant qubit registers and applied during calculation, are the most promising path toward practical quantum computation, several classical post-processing methods have already proven successful in mitigating error.
Errors modeled by bit-flips in the measurement process (``readout error'') can be mitigated in part or entirely, at the cost of additional calibration circuits\cite{Karalekas_2020}.
Errors modeled by a dephasing channel as each unitary operation is applied (``gate error'') tend to result in systematic distortions of the energy surface, as we have seen in the optimization results of Fig.~\ref{fig:Po:RAW}.
These distortions can be mitigated by applying Zero-Noise Extrapolation (ZNE)\cite{Giurgica_Tiron_2020} in which the same measurements are repeated several times, each time modifying the circuit to incur {\em more} noise.
These measurements can then be extrapolated to a hypothetical circuit with zero noise, using Richardson extrapolation or a similar method.

Fig.~\ref{fig:Po:CAL} shows our results on a noisy simulator, applying readout calibration and ZNE for each energy evaluation during the optimization.
ZNE offers noticeable improvement in the highest and lowest bands (calculated independently), but appears less impactful on the intermediate bands (calculated after deflation), perhaps even {\em increasing} variance in the third band.
This may be explained by noting that ZNE is designed to assuage {\em systematic} error, and this is what we tend to observe when we can rely on the variatonal principle, where energies cannot in principle be measured below the ground-state energy.
This is not {\em always} true because our energy estimates are linear combinations of stochastically evaluated Pauli expectation values, and on occasion we do observe trials which appear above the highest band, but these points are relatively rare, and the average values on the highest and lowest bands are shifted inwards.
However, the deflation circuits $\hat V_0$ are somewhat different for each trial, depending on exactly what eigenstate was selected for the lowest band, and this, coupled with the coherent qubit error, has the effect of inducing a {\em random} noise on the intermediate bands.
This explains why the intermediate bands seem to suffer a larger variance but reasonable average values.
We note that many other error mitigation techniques besides readout calibration and ZNE have been proposed in the literature, and our results can likely be improved greatly by implementing more of them.
Nevertheless, the best solution to combat random error remains averaging over more and more trials.

In addition to statistics from a calibrated simulator, Fig.~\ref{fig:Po:CAL} also shows data from the IBM devices \verb|ibmq_athens| and \verb|ibmq_santiago|.
These are calibrated energy measurements of the eigenstates given by the least-error optimization runs on the (calibrated) noisy simulator.
Results are generally consistent with the simulator, but our error mitigation is evidently even less effective on the real devices, and we note that \verb|ibmq_santiago| is especially ill-behaved at certain points.
This may be due to less effective thermal isolation from its environment at the hardware level.
Finally, implementing the controlled-unitary operations necessary for the QPE procedure on a linear architecture introduces an overwhelming amount of overhead in the form of additional SWAP gates, making the QPE refinement part of our algorithm completely intractable on these devices.

\section{Discussion}
\label{sec:Discussion}

We have presented an application of VQD to calculate the band structure of a periodic system.
This algorithm is hypothetically successful in producing accurate results on a device with low noise and is functional to a limited extent on current NISQ devices.
In this section, we carefully analyze the complexity of the algorithm.
The classical approach to band structure includes up to Eq~\ref{eq:H:matrix}, at which point the calculated values $H_{\alpha\beta}(\vb{k})$ are arranged into a Hermitian matrix.
This matrix can be diagonalized using row-reduction or a similar technique in \bigq{M^3} steps, where $M$ is the number of atomic orbitals per unit cell.
This is the standard against which we must compare our quantum algorithm.

Quantum resources are employed in the VQD phase of our algorithm during the operator estimation procedure, for every evaluation of the energy $E \equiv \expval*{\hat H_{\vb{k}}}$.
Each application of the ansatz from Fig.~\ref{fig:ansatz} requires $M$ qubits, \bigq{M} entangling gates, and has a depth between \bigq{\log M} and \bigq{M} layers, depending on qubit architecture.
The Hamiltonian in Eq.~\ref{eq:H:Pauli} has \bigq{M} commuting groups, even including additional terms from the deflation procedure (Eq.~\ref{eq:H:deflation}).
Our implementation requires an ensemble size of \bigo{\epsilon^{-2}} for each commuting group in $\hat H$ to obtain an expectation value accurate within $\epsilon$, but since $\epsilon$ does not scale with $M$, we omit it in the present analysis.
The ensemble states may be prepared sequentially, for a worst-case (linear architecture) execution time on the order of \bigq{M^2}.
Alternatively, the ensemble states may be prepared in parallel, decreasing execution time at the cost of additional qubits.
In the best case, implementing ``classical shadowing''\cite{Huang_2020} reduces the number of required measurements to \bigq{\log M}, and a fully-connected architecture permits a circuit depth as low as \bigq{\log M}, bringing our algorithm into a sub-polynomial quantum resource requirement.
However, the operator estimation procedure is still bounded by the number of Pauli words \bigq{M^2} when measurement results are assembled into the energy $E(\vb*{\theta})=\sum_i a_i \expval{\hat P_i}{\Psi(\vb*{\theta})}$.

Operator estimation is repeated for each function evaluation in the optimization procedure.
The number of function evaluations required depends on the optimization routine selected and the shape of the energy surface, so it is difficult to estimate.
Quantum variational algorithms are notoriously vulnerable to so-called barren plateaus, regions in the parameter surface with a vanishing gradient expected to result in an exponential number of function evaluations for successful convergence.\cite{McClean_2018,Cerezo_2021}
However, research into barren plateaus has focused mainly on densely-packed ansatze which alternate between parameterized rotation gates and entanglement among {\em each qubit}.
The ansatz we have presented has a more constrained structure which rotates and entangles only two qubits at a time.
Additionally, Cerezo et al.\cite{Cerezo_2021} found that local cost functions such as the Hamiltonian we have employed are much more resistant to barren plateaus compared to global cost functions.
Therefore, we conjecture that the number of function evaluations required in our algorithm may be expected to scale polynomially with the number of ansatz parameters, in our case \bigq{M}, provided sufficient error mitigation to suppress noise-induced barren plateaus, which are independent of the ansatz\cite{Wang_2021}.
Thus, we include a factor of \bigq{M^c}, where $c\ge 1$ depends on the optimization.
The optimization is repeated for each of $M$ energy levels; therefore, the VQD phase of our algorithm has a total run-time on the order of \bigq{M^{3+c}}.

An optional QPE phase may be implemented to estimate the eigenvalue to an arbitrary binary precision $t$.\cite{Dob_ek_2007}
The implementation of QPE we have used requires $M+1$ qubits and \bigw{t} rounds of measurement (see Dob\v{s}\'{i}\v{c}ek et al.\cite{Dob_ek_2007} for a tighter bound).
Each round applies a quantum circuit approximating a unitary operator $\hat U_j=\exp(i\hat H\tau_j)$.
Each time slice in the Suzuki-Trotter expansion of $\hat U_j$ on a linear architecture requires an entangling gate count of \bigq{M^3} and a circuit depth of \bigq{M^2}.\cite{Kivlichan_2018}
QPE is repeated for each of $M$ energy levels, setting the best case run-time of the QPE phase of our algorithm on the order of \bigq{M^3}.
Note also that the simulation time $\tau_j$ scales exponentially with the accuracy of the phase estimation procedure, and the number of time-slices must scale accordingly to maintain an accurate $\hat U_j$.
Thus, QPE tends to incur too much overhead for practical application on present-day NISQ devices.

Altogether, evaluating the band energies for each momentum $\vb{k}$ requires \bigw{M^3} time steps, comparable to the classical approach.
Even with a ``perfect'' optimizer in which the optimal parameters $\vb*{\theta_l}$ are produced instantly ($c=0$), the complexities of operator estimation and QPE alone exhibit the same scaling as classical diagonalization and incur significantly greater overhead from the finite accuracy $\epsilon$.
While in this form band structure calculations are not a strong candidate for quantum advantage, quantum computers {\em are} expected to provide a superior edge when including electron correlation terms such as $t_{\alpha\beta\gamma\delta} c^\dagger_\alpha c^\dagger_\beta c_\gamma c_\delta$ in the Hamiltonian, which introduce factors of exponential complexity in the classical approach.
However, such terms also appear to force us to abandon several simplifications we have made.
First, transforming into reciprocal space no longer enables $H$ to be separated into subsytems of size $M$, meaning many more qubits are required to accurately simulate a periodic system.
Second, considering multiple electrons forces us to adopt a qubit mapping which enforces fermionic antisymmetry, greatly increasing the number of commuting groups in the Hamiltonian.
Finally, our ansatz dimension, entangling gates, and circuit depth can no longer remain linear in the number of qubits while simultaneously remaining robust.
Our hope is that this work will inspire similar simplifications to those we have made here, while remaining applicable to highly-correlated systems.

\section{Conclusion}
\label{sec:Conclusion}
In this work, we have presented a systematic algorithm for evaluating band structures on a quantum computer.
We have demonstrated the viability of implementing this algorithm in noiseless qubit systems, and we have demonstrated several of the difficulties faced when implementing it on present-day NISQ devices.
Given the analogy to the classical band structure problem, our algorithm evidently generalizes to solving the eigenvalues of any Hermitian matrix.
Finally we have demonstrated how state-of-the-art quantum algorithms can be applied with drastically lower resource requirements to correlation-free crystalline systems and motivated similar approaches for highly-correlated systems less accessible to classical computing.

\section*{Conflicts of interest}
There are no conflicts to declare.

\section*{Acknowledgements}
We thank Oliviero Andreussi, Itay Hen, Rosa Di Felice, Marco Fornari, Ilaria Siloi, Virginia Carnevali, and Anooja Jayaraj for useful discussions.
We acknowledge support from the U.S. Department of Energy through the grant \textit{Q4Q: Quantum Computation for Quantum Prediction of Materials and Molecular Properties} (DE-SC0019432).
We are also grateful to IBM for providing quantum computing hardware and software.

%%%END OF MAIN TEXT%%%



%%%REFERENCES%%%
\bibliography{ms}

\end{document}